\documentclass[twocolumn,aps,superscriptaddress]{revtex4-1}
\pdfoutput=1

\usepackage[pdftex]{graphicx}
\usepackage{amsmath}
\usepackage{hyperref}
\usepackage{mathtools}

\DeclarePairedDelimiterX\braket[2]{\langle}{\rangle}{#1 \delimsize\vert #2}

\begin{document}

\title{A new non-iterative self-referencing interferometer in optical phase imaging and holographic microscopy, HOLOCAM}

\author{Martin Berz} 
\affiliation{IFE Institut f\"{u}r Forschung und Entwicklung, 81675 Munich, Trogerstr. 38, Germany, martin.berz@ife-project.com}

\author{Cordelia Berz} 
\affiliation{IFE Institut f\"{u}r Forschung und Entwicklung, 81675 Munich, Trogerstr. 38, Germany, martin.berz@ife-project.com}

\date{\today}

\begin{abstract}
  Phase retrieval and imaging phase measurements are fields of intense
  research. It has recently been shown that phase retrieval from
  self-referencing interferograms (SRI) can be reformulated leading to
  a stable, linear equation provided the amplitude of the field is
  known from prior measurement steps (HOLOCAM).  Consequently, the
  numerical solution thereof is straightforward. This is a big
  achievement since convergence is otherwise not always
  guaranteed. Applications are expected in X-ray microscopy, general
  phase retrieval, holography, tomography and optical imaging.
\end{abstract}

\maketitle

\section{Introduction}
\label{introduction}

Measuring the phase of the electrical field is of primary importance
in many areas such as X-ray microscopy, holographic (phase) imaging
and tomography. A vast range of methods is currently
discussed~\cite{wolf}\cite{picart}\cite{falldorf}\cite{kemper}\cite{rhoadarmer}
which proves that the problem is of interest but has not been solved
to satisfaction yet.

We will discuss possible applications of the self-referencing linear
phase method~\cite{berz} (HOLOCAM) which tackles current
difficulties. The approach is a self-referencing interferometric (SRI)
method. Instead of using a reference beam, it employs prior knowledge
in the intensity which we do have. On that basis, the problem can be
exactly reformulated as a linear equation in the unknown complex
electric field. It has been proven that this method is stable for
fields which have arbitrary pixel values in amplitude and
phase~\cite{berz}. No stabilizing or smoothing terms are needed. The
complex field is obtained by solving a linear matrix equation. The
method can be applied to a wide range of 2D and 3D interferometers
including lenses, gratings, diffractive optical elements or
mirrors. Even interferometers with more than two beams can be used.

\begin{figure}
\begin{center}
\includegraphics[scale=0.35]{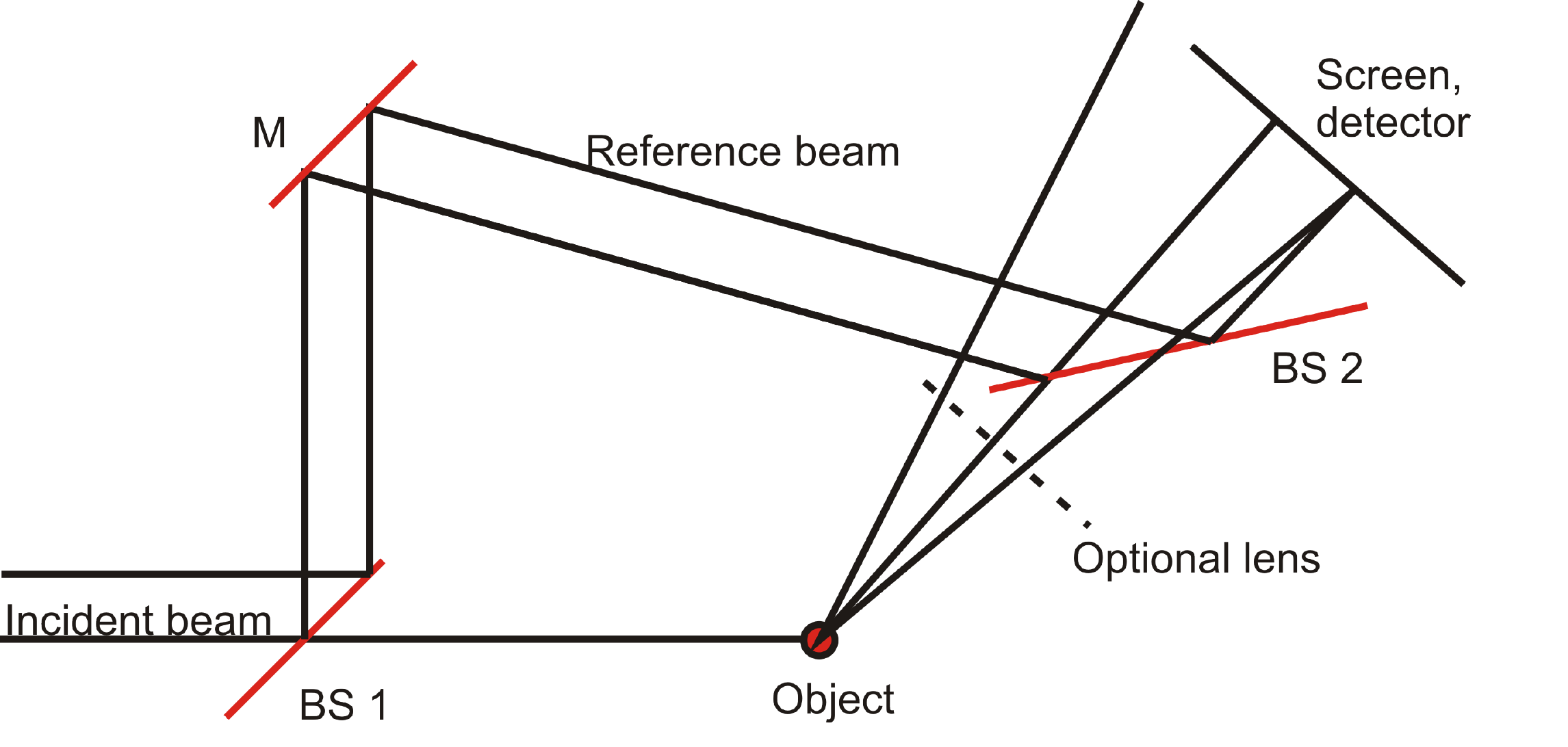}
\caption{\label{reference_beam} The classical holographic setup
  consisting of two beams: an object beam and a reference beam. BS1
  and BS2 are two beam splitters whereof BS1 generates the reference
  beam. M is a mirror which guides the light back to the
  detector. Moving the detector implies that M has to be
  readjusted. Thus, the reference base changes which makes stitching of
  phase images unreliable. This setup has serious drawbacks compared
  to a setup without reference beam such as in Figure
  \ref{general_holocam}}
\end{center}
\end{figure}

In most digital holographic applications the phase is recovered by
interference with a reference beam~\cite{picart}. Although this is a
perfect solution mathematically a reference beam must be available.
In practice, the light from the illumination (usually a laser) is split
into two beams, whereof one beam becomes the reference and the second beam
illuminates the object (Fig. \ref{reference_beam}).

The setup in Figure \ref{reference_beam} shows an interferometer with
a reference beam. Since the object is part of one of the branches
every change of the object, the illumination of the object or the
location of the detector actually leads to a path change in the
interferometer. Hence, it is sensitive to small path differences up to
fractions of a wavelength. Thermal drifts and air fluctuations
generate incoherent noise.  Dust, scratches, defects produce further
coherent noise which is system inherent. It is known that this limits
the accuracy of the phase measurement, i.e. it blurs the
image~\cite{kemper}. Many remedies have been developed such as active
compensation~\cite{marquet} and short exposure times~\cite{yosh}. Yet
this is encompassed by further unwanted physical complexity of the
system.

The final results cannot be better than the initial data
quality. Therefore, attempts are made to get rid of the reference
beam. This is particularly addressed in former work on
self-referencing interferometry
(SRI)~\cite{falldorf}\cite{kemper}\cite{rhoadarmer}. For SRI, it is
vital to disentangle the two fields in the interference signal.  For
this purpose, a mode-filter synthesized reference
beam~\cite{rhoadarmer} or a non-linear functional in a lateral
shearing type interferometer~\cite{falldorf} is used. This is
restricted to special applications though.

The HOLOCAM has advantages compared to other known SRI methods:
firstly the methodology is very simple and secondly it can be applied to
many different interferometers not only to lateral shearing
interferometers.

\begin{figure}
\begin{center}
\includegraphics[scale=0.35]{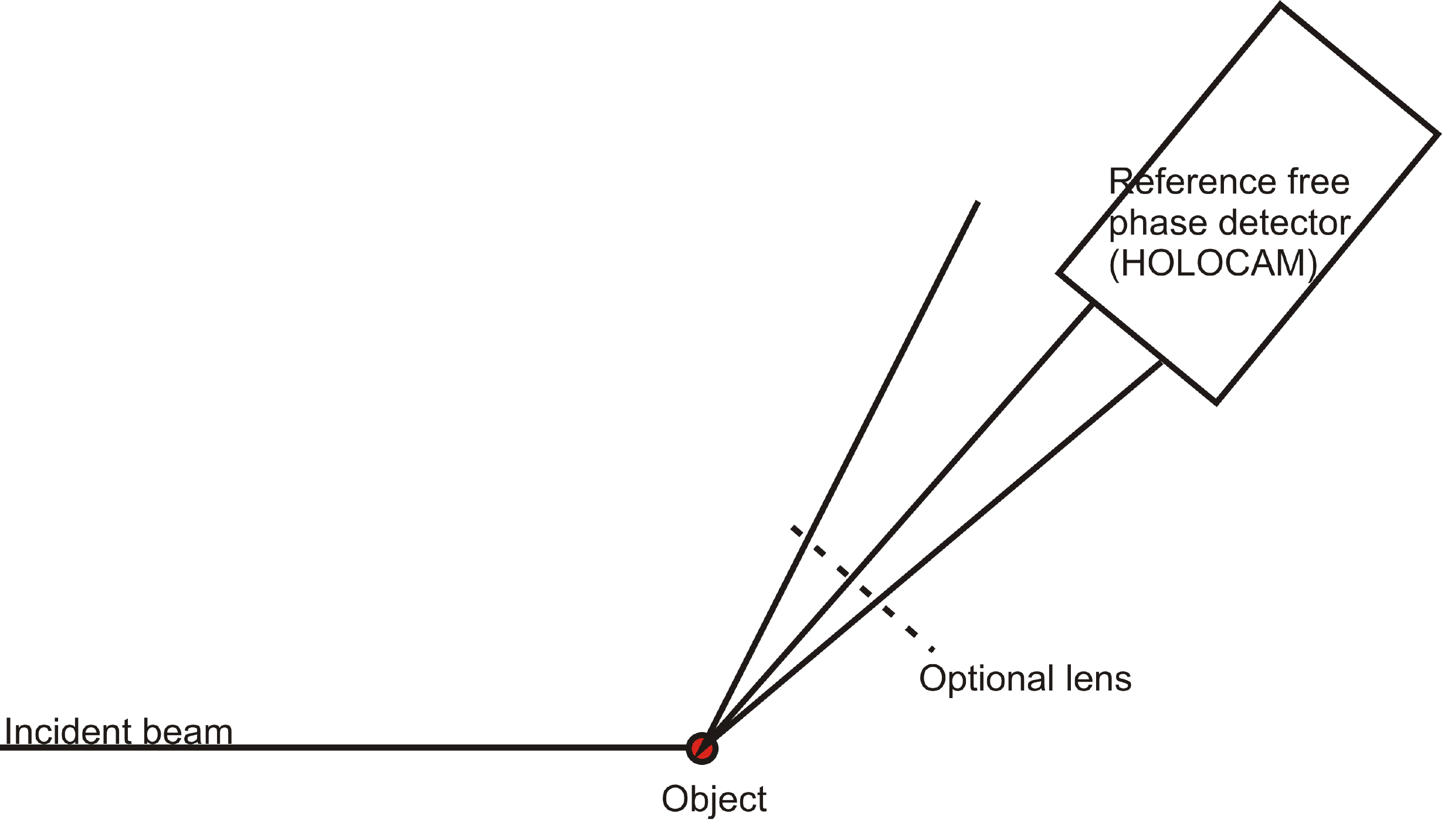}
\caption{\label{general_holocam}A general HOLOCAM setup. The lens is
  needed since the HOLOCAM has a limited field of view. }
\end{center}
\end{figure}

Lastly, the HOLOCAM can be designed as a compact detector box that
measures the phase (Fig. \ref{general_holocam}). All interferometric
parts of the system can be sealed in the detector housing. HOLOCAM is
an artificial word created from HOLOgraphy and CAMera. The device is
holographic since it uses a holographic type evaluation of holograms
or interferograms and it measures something 'holographic', the complex
amplitude. The device is also a camera such as a CCD camera which
registers the locally resolved light intensity on a flat detector. The
HOLOCAM does something similar but the registered quantity is the
complex amplitude field.  Both the CCD and the HOLOCAM do not need a
reference beam.

\begin{figure}
\begin{center}
\includegraphics[scale=0.35]{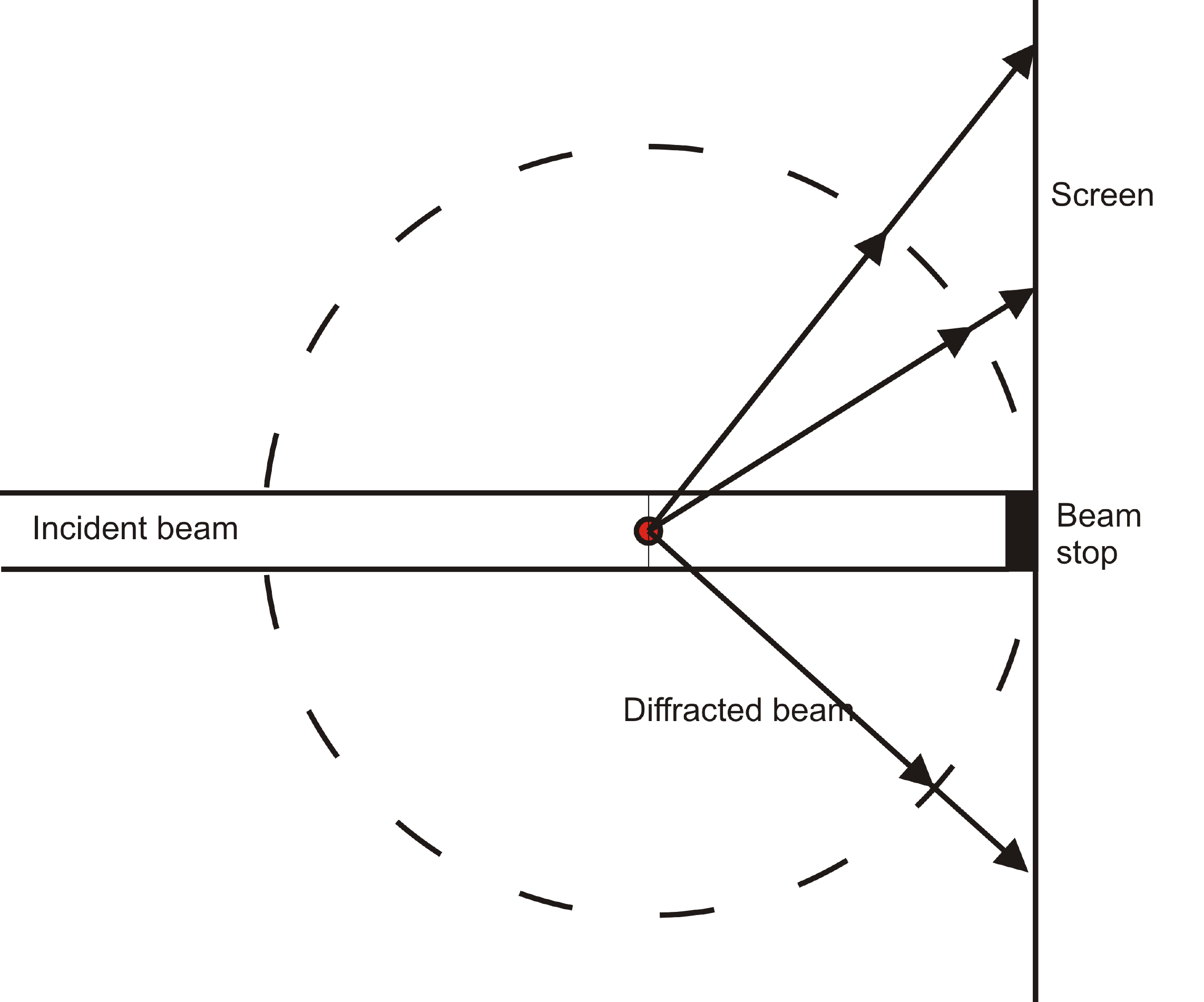}
\caption{\label{x_ray}A classical X-ray scattering experiment.}
\end{center}
\end{figure}

\begin{figure}
\begin{center}
\includegraphics[scale=0.35]{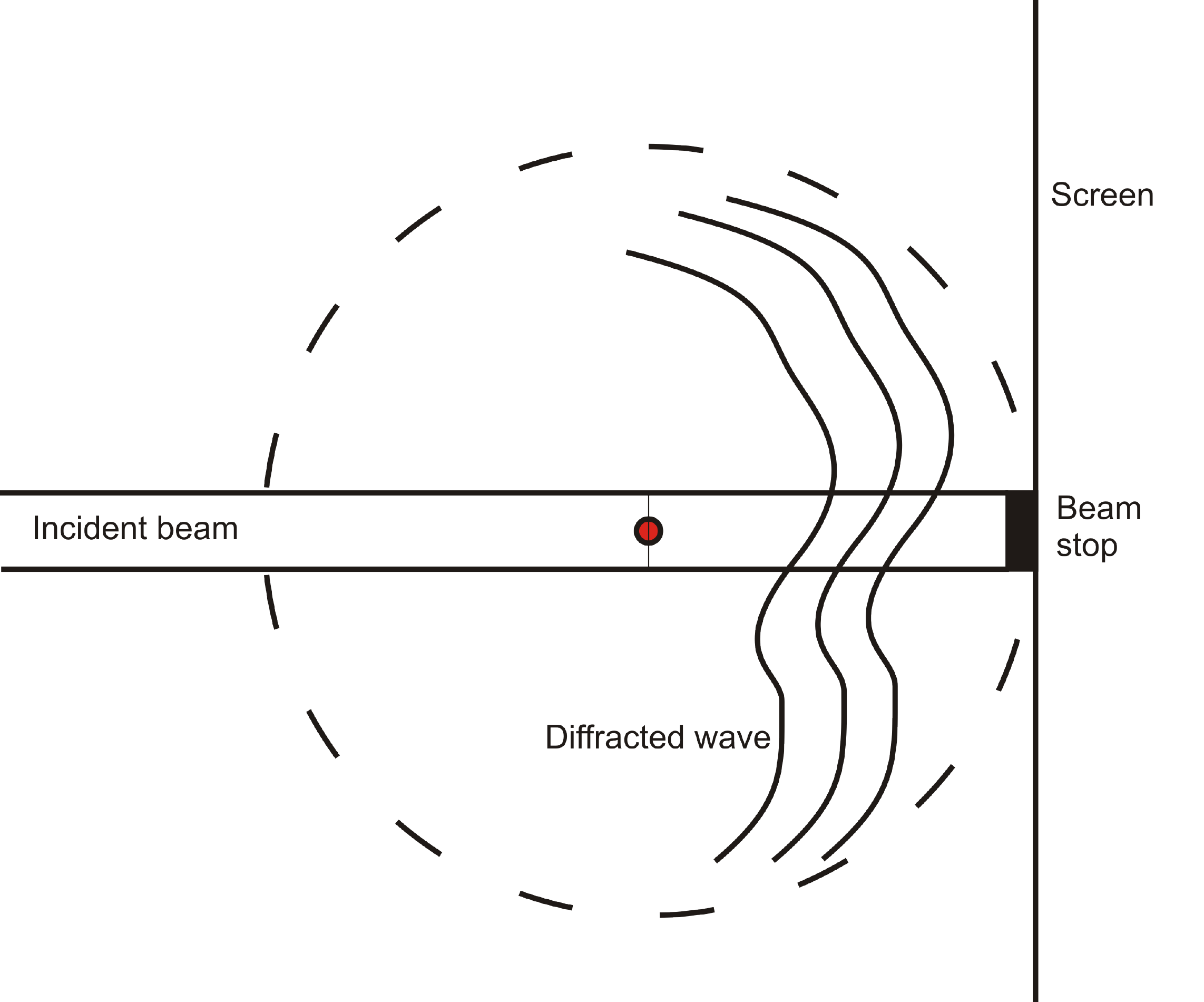}
\caption{\label{x_ray_wave}A X-ray wave emitted in a  scattering experiment.}
\end{center}
\end{figure}

The HOLOCAM can also make contributions to X-ray tomography.
In the early days of physics, X-ray probing was used as a lensless
system (Fig.\ref{x_ray}). This is the honourful work of Laue, Debye and
Scherrer. It is characterized by the measurement of diffraction
intensities (Bragg reflections) on a screen. For real samples (in
particular non-crystalline samples) the scattering intensities are
more complicated including also light intensities besides the distinct
Bragg peaks ('diffraction peaks midway between Bragg peaks'). The
general view of a X-ray scattering process is a scattered wave emitted
from the scatterer (Fig.\ref{x_ray_wave}). Like any other optical
field, this wave possesses an intensity and a phase.

The purpose of X-ray scattering is the determination of the atomic
structure of the scatterer ('the object', 'the sample'). The work on
this exciting topic has led to discoveries such as the structure
determination of the DNA~\cite{watson}, rewarded by the Nobel prize.

The problem is a so called 'inverse problem': the scattering structure
is to be determined, knowing the intensities and the scattering
process, but not the phase of the field~\cite{jaga}. The problem could
be directly solved if not only the intensities but also the phases of
the scattered field were known. This is not the case though. In fact,
some additional knowledge is necessary. Different solution methods can
be distinguished by the kind of prior knowledge needed~\cite{jaga}. It
is indeed challenging to generate 'good' additional experimental
knowledge ('experimental data'). No refractive materials as known from
visible optics exist in the spectral X-ray region. Mirrors only work
at grazing incidence. Until recently, it was not possible to produce
useful X-ray lenses. Even though zone plates~\cite{pearson} are
available novadays it is still very difficult to do imaging with
X-rays. Hence, the principal goal has remained unchanged: determine
the structure of the object from intensity data of the scattering
process. This is called 'X-ray imaging' since imaging is the purpose
of these experiments.

\begin{figure}
\begin{center}
\includegraphics[scale=0.35]{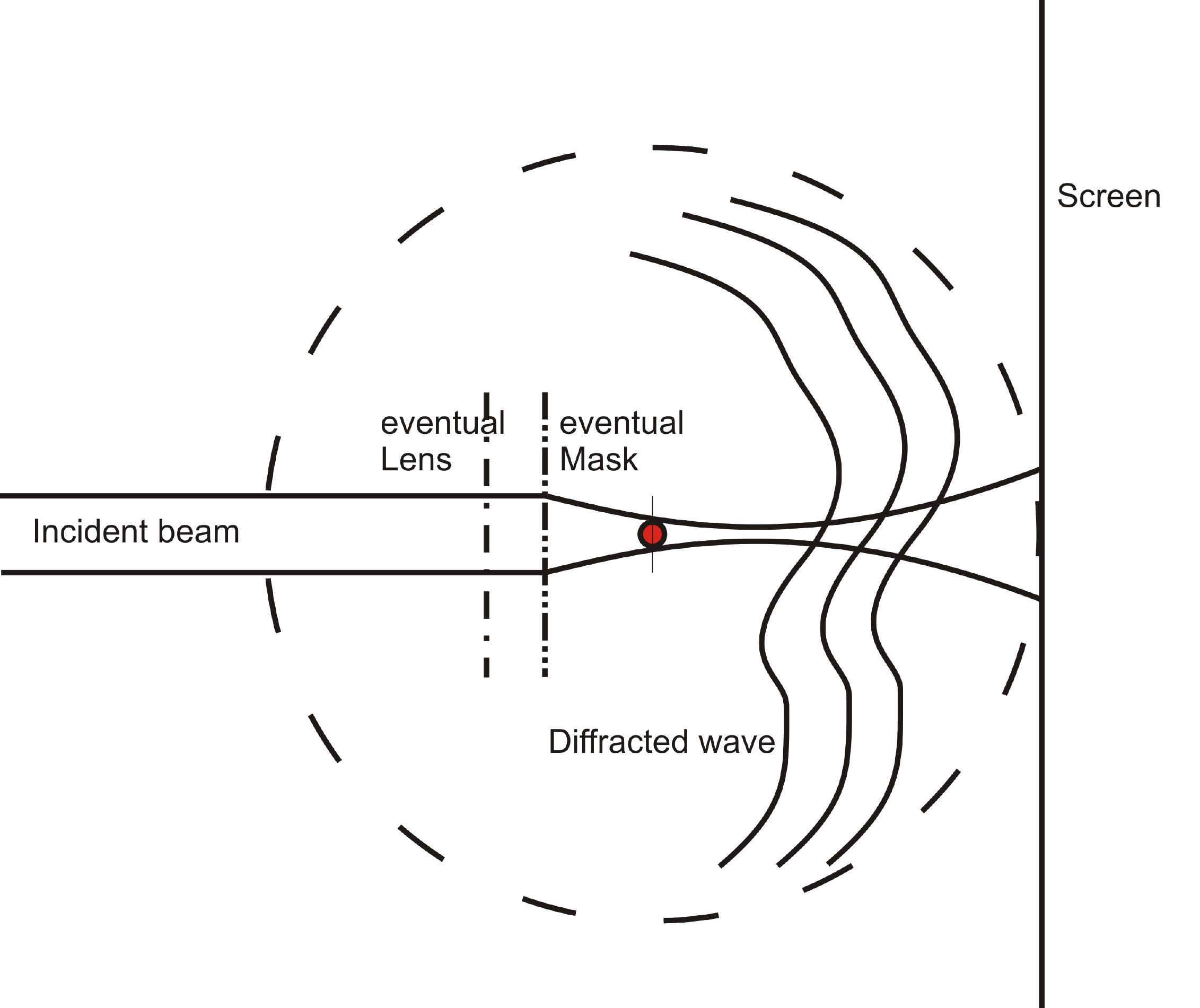}
\caption{\label{add_infox_ray} Generation of additional information in
  a X-ray imaging experiment. Exposures with different masks or
  different object-lens positions are necessary to provide prior knowledge for
  the intensity based phase retrieval process (Section \ref{theory})}
\end{center}
\end{figure}

In general, the needed prior knowledge has to be generated by
additional measurements. One type of useful manipulation is shown in
Figure \ref{add_infox_ray}. In many cases no lens is present between
the object and the detector (so called 'lens-less imaging'). Thus, a
possible approach is to vary the field that is incident on the object
using either masks or a lens before the object
(Fig. \ref{add_infox_ray})\cite{chapman}.

In conclusion, quite some effort is necessary to generate the
additional information vital for the reconstruction process. In
addition to that, the reconstruction process should work independently
of the scattering object, at best with little or no adjustments. The
reconstruction process is usually based on numerically intensive,
iterative steps. Only good data lead to a sharp object image.

\begin{figure}
\begin{center}
\includegraphics[scale=0.35]{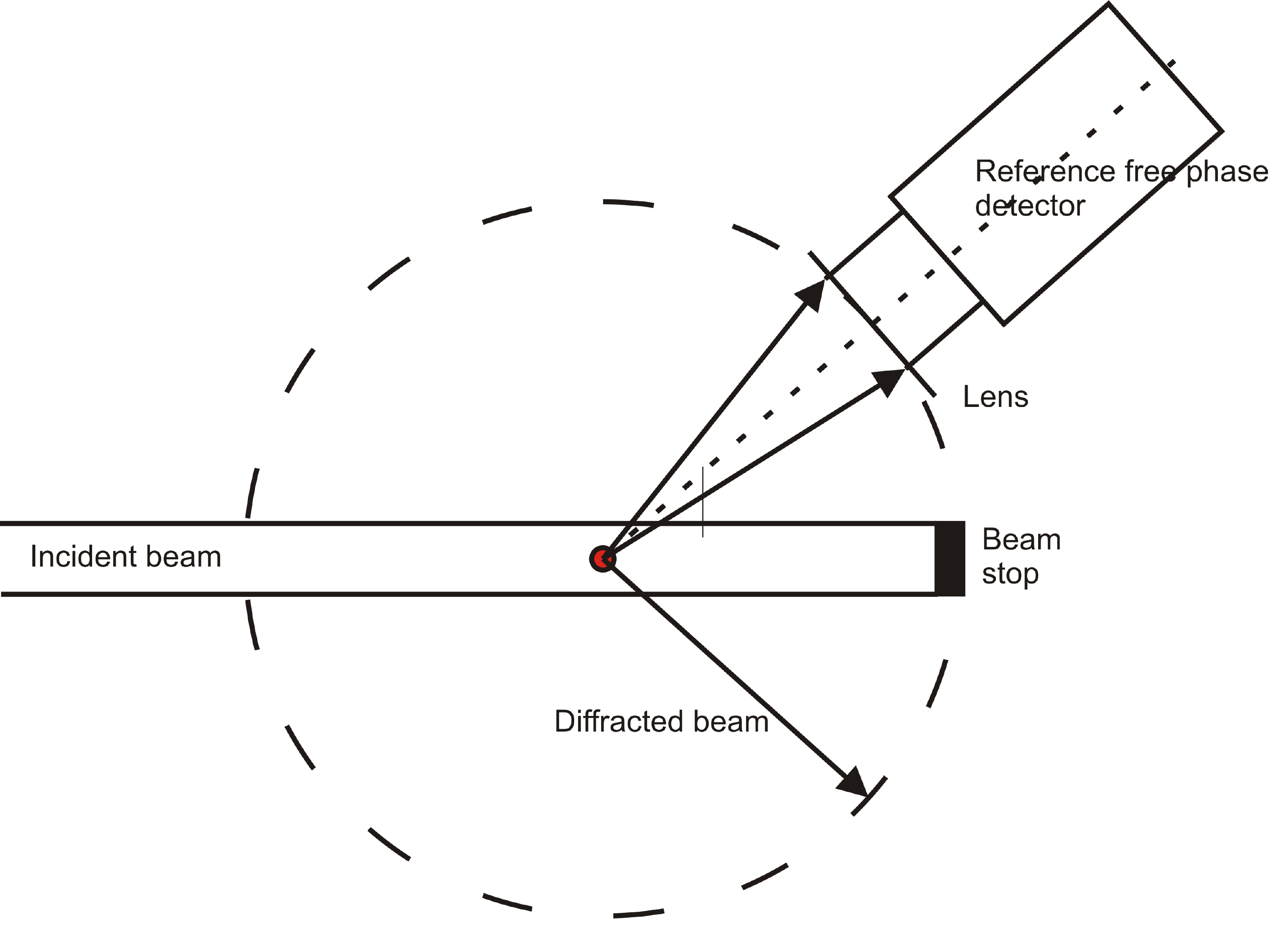}
\caption{\label{x_ray_holocam} Use of a HOLOCAM in a X-ray scattering
  experiment. Aberrations at the lens in front of the HOLOCAM detector
  can be corrected numerically. The detector can be displaced to different
  locations ('stitching of phase information') The evaluation of the
  phase information allows for a resolution at the diffraction limit.}
\end{center}
\end{figure}

\begin{figure}
\begin{center}
\includegraphics[scale=0.35]{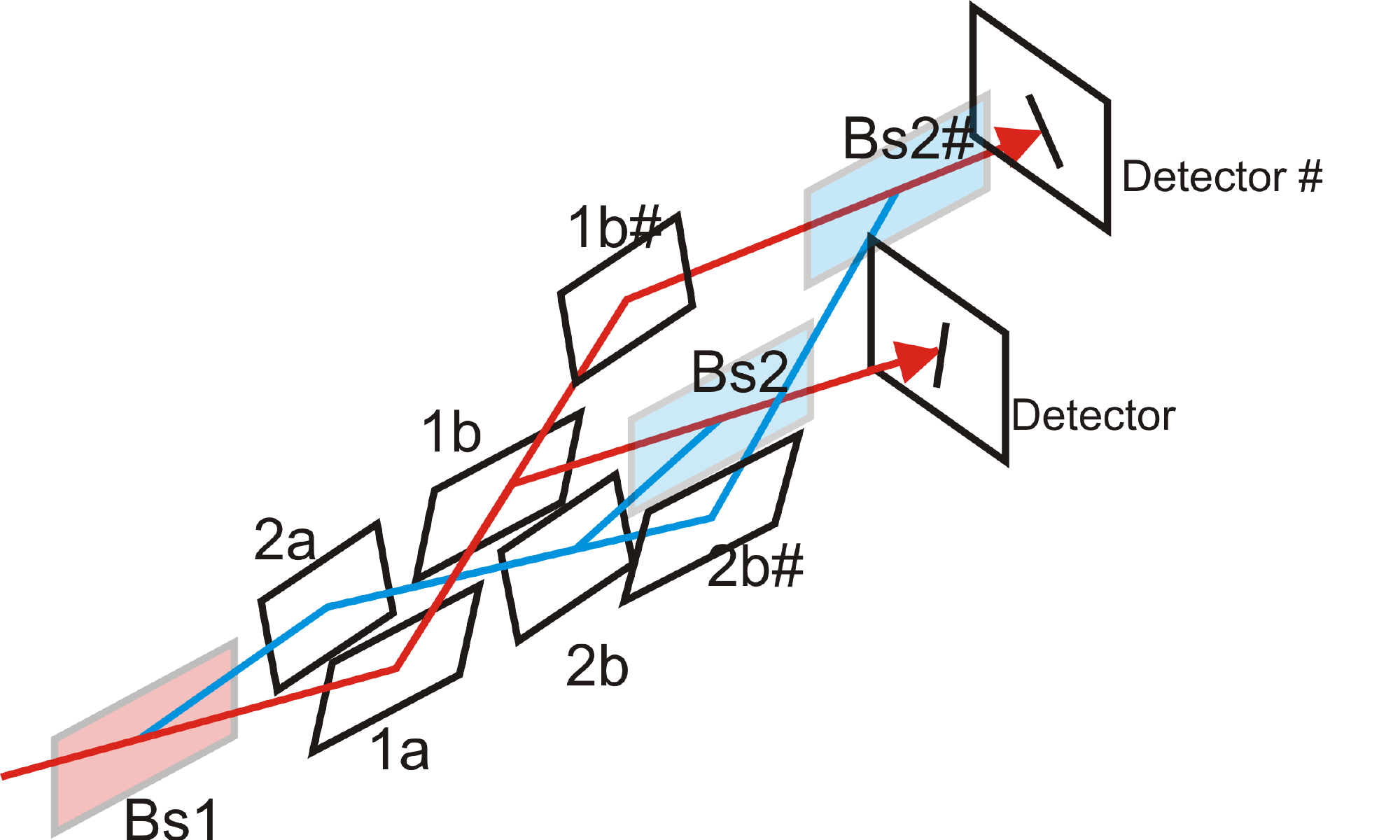}
\caption{\label{x_ray_gracing_holocam}Design of a HOLOCAM for X-rays,
  using grazing incidence mirrors. The gracing incidence mirrors 1a,1b
  and 2a,2b form a kind of wave guide which mutually turns the
  fields. Two images are needed for uniqueness which are distinguished
  by the '\#' sign.}
\end{center}
\end{figure}

That is where the HOLOCAM could come into play. Figure
\ref{x_ray_holocam} shows a possible setup. It can be seen that the
HOLOCAM needs a lens such as in any other common optical
arrangement. This is certainly a drawback compared to lensless systems
but there are two advantages which should compensate for the higher
effort caused by a lens: For a HOLOCAM system, lens aberrations can be
corrected numerically and the HOLOCAM gives directly non-iterative
phase data. Lenses with sufficient quality are therefore
available~\cite{pearson}. Hence, the potential merit of the HOLOCAM is
that it allows to improve the tomographic resolution up to the
diffraction limit even if the lens is imperfect. The HOLOCAM can be
built internally with grazing incidence mirrors
(Fig. \ref{x_ray_gracing_holocam}).

In Section \ref{theory}, we will give a brief summary of the theory
involved. More details can be found in Reference \cite{berz}.

In Section \ref{holocam_visible}, we will discuss some more aspects of
  the HOLOCAM applications in the visible spectrum. In section
  \ref{holocam_x_ray} this will be continued for X-rays.

\section{The phase measurements}
\label{theory}
\subsection{The intensity based phase retrieval}
We start by outlining the basics of phase retrieval. The light field
is a complex field consisting of intensity and phase
information. Physically, only intensities can be measured
directly whereas phases must be determined indirectly. This is the origin
of the phase retrieval problem, i.e. the task to determine the complex field
$f$ from its intensity (or 'amplitude') $\mid f \mid$. Of course,
stated that way, the problem is undetermined. The problem becomes
tractable if some prior knowledge is available. Consequently, phase
retrieval problems can be distinguished by the kind of prior knowledge
used.

In optics, the scattering problem is of great importance since it
allows the optical inspection of a specimen such as the structure of
DNA~\cite{watson}. In this case, the intensities of the Fourier
coefficients are given. However, this does not improve the general situation
as it merely reformulates the problem: Determine $f(x)$, a
complex quantity, knowing $|F(k)|$:

\begin{equation}
\label{eq:basic_IF}
F(k)=|F(k)|e^{{i\psi (k)}}=\int _{{-\infty }}^{{\infty }}f(x)\ e^{{-2\pi ik\cdot x}}\,dx
\end{equation}

The Fourier transformation is an injective mapping which immediately
proves that this problem cannot have a unique solution except
for cases of further prior knowledge. In the case of the Gerchberg and
Saxton (GS) algorithm, this prior knowledge consists of two (instead of one)
intensity measurements such as $|F(k)|$ and
$|f(x)|$~\cite{gerchberg}. Alternatively, the intensity can be measured
with masks before or after the object (structured illumination,
oversampling and
sparsity)~\cite{jaga}\cite{chapman}\cite{miao}\cite{dremeau}.

Phase retrieval techniques are collectively called 'intensity based
phase retrieval'. The method has the following characteristics: No
reference beam is needed but additional measurements (under different
physical conditions). It is an iterative method and numerically
very demanding. 

\subsection{The reference based phase retrieval}

In this method, the interference $IN_s$ between the light field and a
reference beam is measured\cite{osh}.

\begin{equation}
\label{eq:basic_IF1}
IN_{s} := \overline {( E1_s + E2_s)} *  {( E1_s + E2_s)} 
\end{equation}

's' denotes some indexation of pixels of the image recording
device. In this case, $E1$ is the field of the unknown phase and $E2$ is a
field of known phase.

\begin{equation}
\label{eq:basic_IF2}
IN_s = \mid E1_S \mid^2 + \mid E2_S \mid^2  + 2 \ Re ( \ \overline{E2_S} * E1_S \ )
\end{equation}

Using known methods like 'phase shifting' or 'carrier
phase'\cite{yosh} the complex term IF is determined:

\begin{equation}
\label{eq:basic_IF3}
IF_{s} := \overline{E2_s} * E1_s 
\end{equation}
  
If $E2_s$ is known $E1_s$ can be calculated directly. The knowledge of
$E2_s$ is a highly demanding type of prior knowledge since it involves
the physical use of a reference beam of known properties.

The methods above will be called 'reference based
phase retrieval'. The method has the following characteristics: A
reference beam is needed with the advantage of a simple numerical
evaluation.

\subsection{The correlation based phase retrieval (or SRI methods)}

Self-referencing interferometer (SRI) methods represent a third
class~\cite{falldorf}\cite{kemper}\cite{rhoadarmer} which
includes all approaches that determine the phase by interferometric
means and that do not use an 'external' reference beam. The spread of
methods is quite large. The SRI approach is also called 'correlation
based phase retrieval'.

One method is to generate a zero mode optical field, which serves as a
synthetic reference beam~\cite{rhoadarmer}. This can be achieved by a
spatial band pass such as an illuminated pinhole or single mode
fiber leading to a loss of light though. Furthermore, the
process  generating the reference involves some intensity arbitrariness of
the fields introducing noise sources in the overall
process. This kind of SRI is thus not a really satisfying solution yet.

Another approach is to measure lateral shearing
interferograms~\cite{falldorf}\cite{kemper}. The phase is obtained by
a iterative optimization of a non-linear functional

\begin{equation}
  \label{eq:sri_functional}
  \left\Vert IF - \overline{E1} * S(E1) \right\Vert \rightarrow 0
\end{equation}

S represents a lateral shear,
either a x shear or a y shear. A convergence ensuring term is added
which smooths the result. In general, convergence to a global minimum
and stability are big challenges in phase reconstruction.

Lateral shearing interferometers do not generate particularly
characteristic interferograms. This can be seen from the fact that the
lateral shearing interferometer essentially measures the derivative of
the phase~ \cite{malacara}. As a consequence, plane waves show up as
constant offsets in the interference signal which can only badly be
resolved. In conclusion, incident optical plane waves cannot be well
detected. Apart from that, it will be shown later in this article, that
lateral shearing interferometers are difficult to calibrate since they
do not have a fixed point. This will be explained in the context of
the HOLOCAM method (Subsection \ref{holocam_theory}).

This kind of SRI approaches are called 'classical SRI' in contrast to
the HOLOCAM method which might be considered as a SRI technique, too.

'Classical SRI' has the following properties: No reference beam is
needed. Depending on the particular approach the resolution of the
method is limited.

\subsection{The HOLOCAM phase measurement}
\label{holocam_theory}

\begin{figure}
\begin{center}
\includegraphics[scale=0.35]{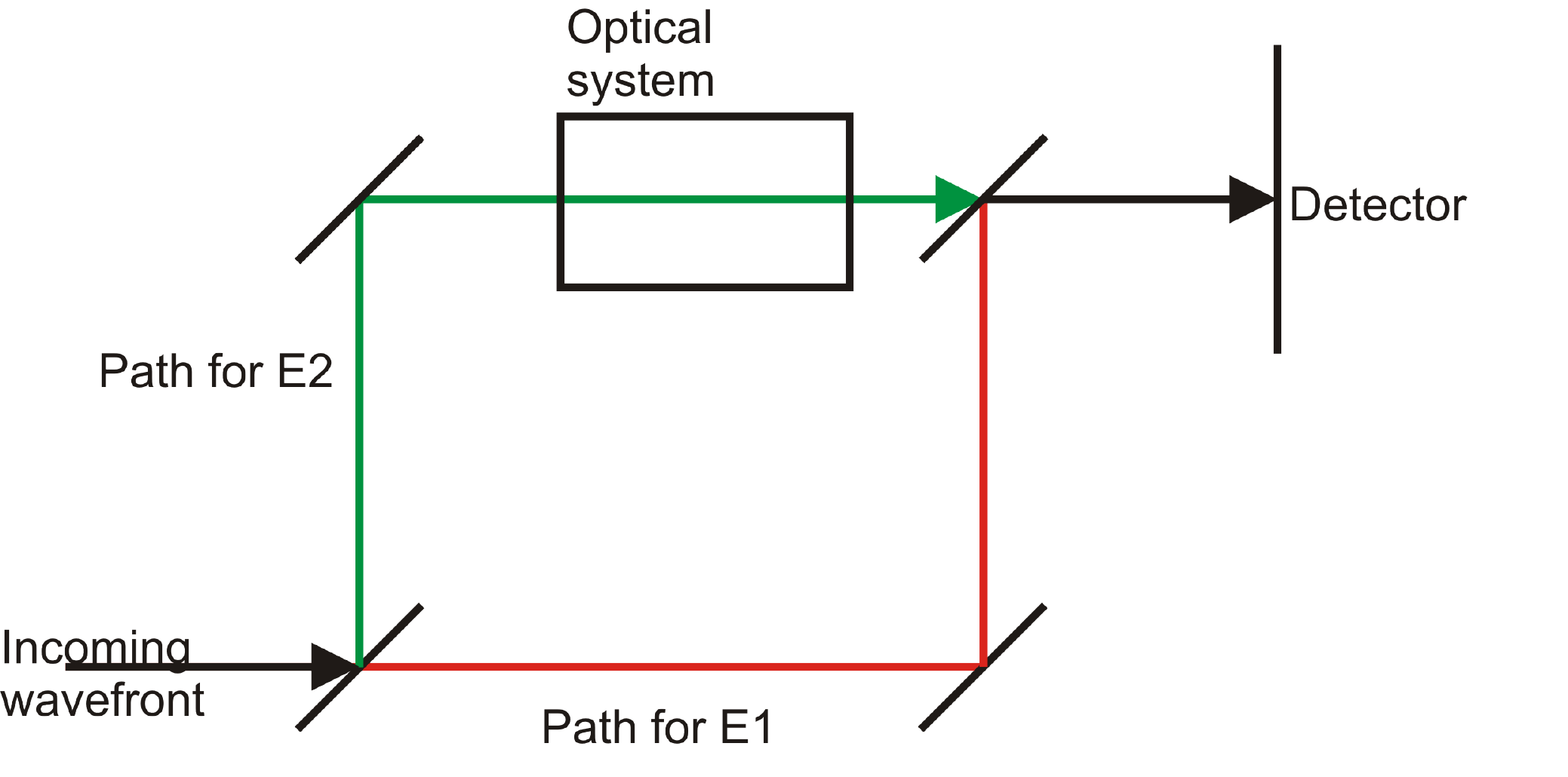}
\caption{\label{gen_holocam}The HOLOCAM principle in general. The field
  propagation for the two branches is different which is symbolized by
  an optical system in the upper branch.}
\end{center}
\end{figure}

Abstractly speaking, the method divides the field E into two parts
$E1$ and $E2$ (Fig. \ref{gen_holocam}). The interferometers used for the HOLOCAM have the
property that $E2$ is a known linear function of $E1$. This assumption
is not very restricting. In some cases the mapping might even be very
simple (Section \ref{holocam_visible} or Reference
\cite{berz}). U is the back propagation of $E1$ by branch to a
plane before the HOLOCAM and subsequent forward propagation of this
field by branch 2 to the detector plane. If the detector plane
mapped by U corresponds to two conjugate planes, the map U becomes a
geometric point mapping. 

\begin{equation}
\label{eq:basic_U}
E2_{s} = \sum_{t} U_{s,t} E1_t =: U(E1) =: U E1 
\end{equation}

Using this the expression for IF can be reformulated.

\begin{equation}
\label{eq:if_1}
IF_s = \overline{E2_{s}} * E1_{s} = E1_{s} * \sum_{t} \overline {U_{s,t}
  E1_t}
\end{equation}

Multiplying this equation with $\overline{E1}$ yields

\begin{equation}
\label{eq:fund_eq}
IF_s * \overline{E1_s} =  \mid E1_s \mid^2  \ * \ \sum_{t} \overline{U_{s,t}} \overline{E1_t} 
\end{equation}

Having prior knowledge in $\mid E1 \mid^2 =: I1$ this becomes a
linear equation for $\overline{E1}$:

\begin{equation}
\label{eq:fund_eq2}
\sum_{t} ( I1_s \overline{U_{s,t}} - IF_s * \delta_{s,t} )  \overline{E1_t} = 0
\end{equation}

$\delta_{s,t}$ is the Kronecker function. Equation (\ref{eq:fund_eq2})
is a linear equation in $\overline{E1}$ since $U$ and $I1$ are
known. This is called the fundamental equation for the HOLOCAM.

The intensity $I1$ can be measured by blocking branch 2 or by an
additional beamsplitter which allows to measure the light $E1$ on a
separate detector. For simplicity this almost trivial task will not be
shown in the figures.

The dimension of the subspace of the zeroth eigenvectors of the
fundamental equation must be checked for each case. For a unique solution the
dimension has to be one. This can be achieved by an appropriately
designed interferometer. Figure (\ref{vis_holocam}) shows an example
which is also discussed in Reference \cite{berz}. Another example is
shown in Figure (\ref{x_ray_gracing_holocam}). It actually represents
a HOLOCAM with two measured correlations.

The solution space of linear equations can be completely
characterized. Hence, there is neither a convergence nor a stability
problem~\cite{berz}.

The stability of the fundamental equation is of great importance. To
some extent this can be verified by analyzing reference \cite{osh}. It
has been recognized by Osherovish et al. that prior knowledge in the
intensities of interfering felds renders the phase determination more
stable. This has been derived for an interference with a reference
beam. Investigations on the HOLOCAM approach have verified that this
is also true for SRI~\cite{berz}. This seems to be a cornerstore for
the stability of the HOLOCAM method in which the intensity of the
field is used as prior knowledge.

The HOLOCAM devices differ in their mapping U. It is favorable to have
a fixed point in U. This entails the existence of a pixel with index
$s$ for which $E1_s$ and $E2_s$ correspond to the same point of the
incoming wave front. An example is the rotation of an electric field
where the center of the rotation is mapped onto itself. At the fixed
point, the interference has a real value in IF. Hence, it can be used
to determine the absolute phase of IF. Therefore, a fixed point in the
mapping U allows an absolute calibration of IF. A pure lateral
shearing interferometer has no fixed point which explains some of the
difficulties with lateral shearing interferometers (Section
\ref{introduction}).

The HOLOCAM approach is quite general. To give an example,
interferograms obtained by lateral shearing interferometers in
'classical' SRI can also be evaluated by the HOLOCAM method. As a
consequence, the linear fundamental equation Eq. (\ref{eq:fund_eq2})
has to be solved using an operator U which is a lateral shear (or
shift). Furthermore, the intensity $|E1_s|^2$ has to be
known. Besides, the solution obtained by this method is the global
minimum of the functional Eq. (\ref{eq:sri_functional}). Without
noise, the solution of the fundamental equation is a pointwise
solution of Eq. (\ref{eq:fund_eq2}) and the functional
Eq. (\ref{eq:sri_functional}) becomes even zero. This must be the
global minimum of the optimization problem Equation
(\ref{eq:sri_functional}). Of course the solution of a linear matrix
is simpler then the search for a global minimum.

A drawback of lateral shearing is the lack of a fixed point
calibration method. Calibration is an important aspect for all
methods, not only for the HOLOCAM. Consequently, interferometers with
fixed points should be used preferentially for the HOLOCAM method.

The HOLOCAM approach is also called 'linear correlation based phase
retrieval'.

In conclusion, phase retrieval can be divided in four categories:

\begin{itemize}
\item Intensity based phase retrieval ('pure' phase retrieval)
\item Reference based phase retrieval ('classical' holography)
\item Correlation based phase retrieval ('classical' SRI)  
\item Linear correlation based phase retrieval (HOLOCAM)
\end{itemize}

\section{HOLOCAM visible imaging}
\label{holocam_visible}

\begin{figure}
\begin{center}
\includegraphics[scale=0.35]{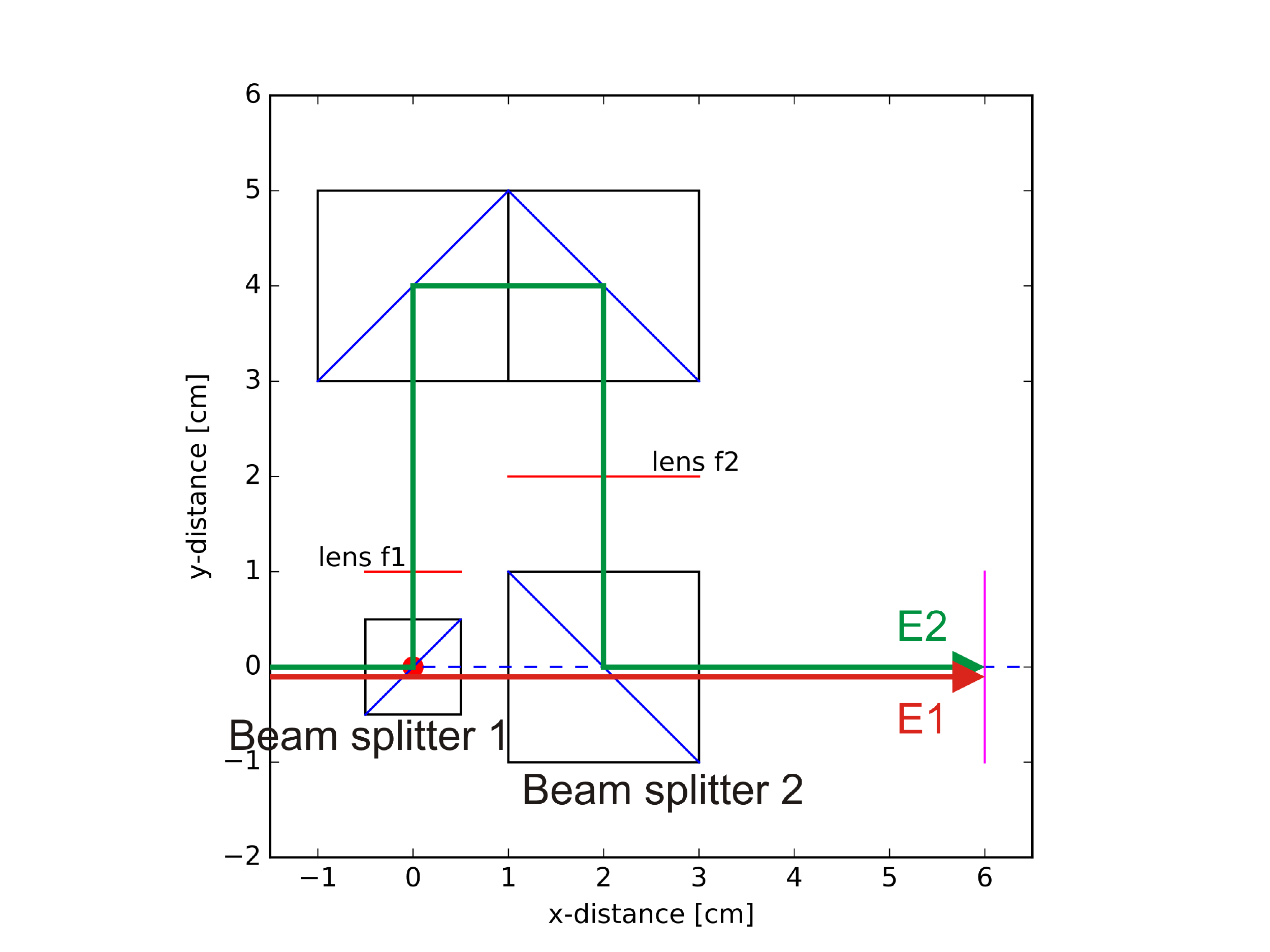}
\caption{\label{vis_holocam}Design example for a HOLOCAM in the
  visible spectrum. The setup is conceptually simple and therefore
  useful as a demonstration. Advanced HOLOCAM systems might exploit
  the large design freedom for compactness. Further more, it is
  possible to choose a HOLOCAM design with equal path lengths. }
\end{center}
\end{figure}

Figure \ref{vis_holocam} shows a demonstration example of a HOLOCAM. A
pair of lenses is inserted in one of the branches of a Mach-Zehnder
interferometer in such a way that U becomes a point mapping. U introduces a
radial shear with one fixed point (actually a beam expander). This
configuration has been extensively investigated by numerical
simulations~\cite{berz}. This simple example is quite useful to get
an understanding of the HOLOCAM concept. The interferometer in Figure
\ref{vis_holocam} has unequal path lengths which is not a necessity
but a particularity of this design.

Another example is a lateral shear interferometer with two
recorded IF images for x and y shear
respectively~\cite{falldorf}\cite{malacara}. This configuration has
no fixed points.

The two branches are equal in path length. It has been shown that $10\mu m$
coherence length of a light emitting diode are sufficient to record
good quality IF images~\cite{falldorf}. The experimental proof was
given for a method minimizing the functional
Eq. \ref{eq:sri_functional}, but is still valid for a HOLOCAM type
evaluation, as shown at the end of Subsection \ref{holocam_theory}.

The HOLOCAM method avoids an iterative solution process and can be
applied to much more general configurations as shown by the 3D
example in Figure \ref{x_ray_gracing_holocam}. Multiple scattering
devices and diffractive optical elements DOE can also be
used.

Next, we will analyze the measurement of the complex interference term
IF. Only real quantities can be measured. This sounds similar to the
initial phase reconstruction problem but is in fact quite a different
question compared to Section \ref{introduction}. IF is, by definition,
equal to $\overline{E2}E1$ which involves the phase of the product of
the two fields (phase difference). The determination of
a phase difference can be tackled by established methods such as
'carrier phase method' and 'phase shifting method'~\cite{yosh}. It is
important to note that this can be done within one recorded
real-valued interferogram ('one shot technique').

We also want to discuss the role of polarization. It can always be
assumed that the field at the entry of the HOLOCAM is linearly
polarized. If this should not be the case it can be enforced by a
polarizer which might even probe both polarizations
consecutively. Additional effects of polarization rotation might exist
inside the HOLOCAM interferometer limiting the degree of
interference. It is always possible to use a further polarizer just
before the detector to ensure 100\% interference though.

Polarization is interesting for another reason. The two light paths of
the HOLOCAM can be designed to have different polarizations. Even if
they occupy the same area in space they can be manipulated separately
by polarization sensitive equipment allowing to design of ultra
compact HOLOCAM devices.

The HOLOCAM strategy is mainly a two step process. In the first step,
the complex phase of $E1$ is determined. This is the proper HOLOCAM
step. In the second step, this information can be used for subsequent
operations such as the propagation of the phase information to another
imaging plane or as a part of a larger data series in a tomographic
analysis. This stepwise approach allows us to monitor the quality within
the process. An example is the size of the lowest eigenvalue of the
fundamental equation Eq. (\ref{eq:fund_eq2}). In an ideal case this
eigenvalue should be zero but error influences such as detector
quantization lead to a finite value~\cite{berz}). The corresponding
eigenvector is still the electrical field $E1$~\cite{berz}. Hence,
every individual step of the HOLOCAM method can be calibrated and
optimized independently. Another advantage is a better control of
subsequent steps. For instance,  the 'twin problem' known from holographic
evaluations does not exist in this approach~\cite{liebling}.

The HOLOCAM can be used in all known
configurations~\cite{picart}\cite{yosh} in holography or diffraction
tomography. Aberrations in optical components can be corrected
numerically. Electronic focusing can be applied\cite{liebling}.

Nowadays, a reference beam based determination of $E1$ is mainly used in
digital holography (Reference based phase retrieval or classical
holography). Besides complexity and noise problems, this has the
disadvantage that the detector cannot be displaced
(Fig. \ref{reference_beam}).  The HOLOCAM system does not have this
limitation. To give an example, this would solve the
known problem of the missing frequencies in tomography~\cite{kostencka}.

\section{HOLOCAM 3D interferometer X-ray imaging}
\label{holocam_x_ray}

\begin{figure}
\begin{center}
\includegraphics[scale=0.35]{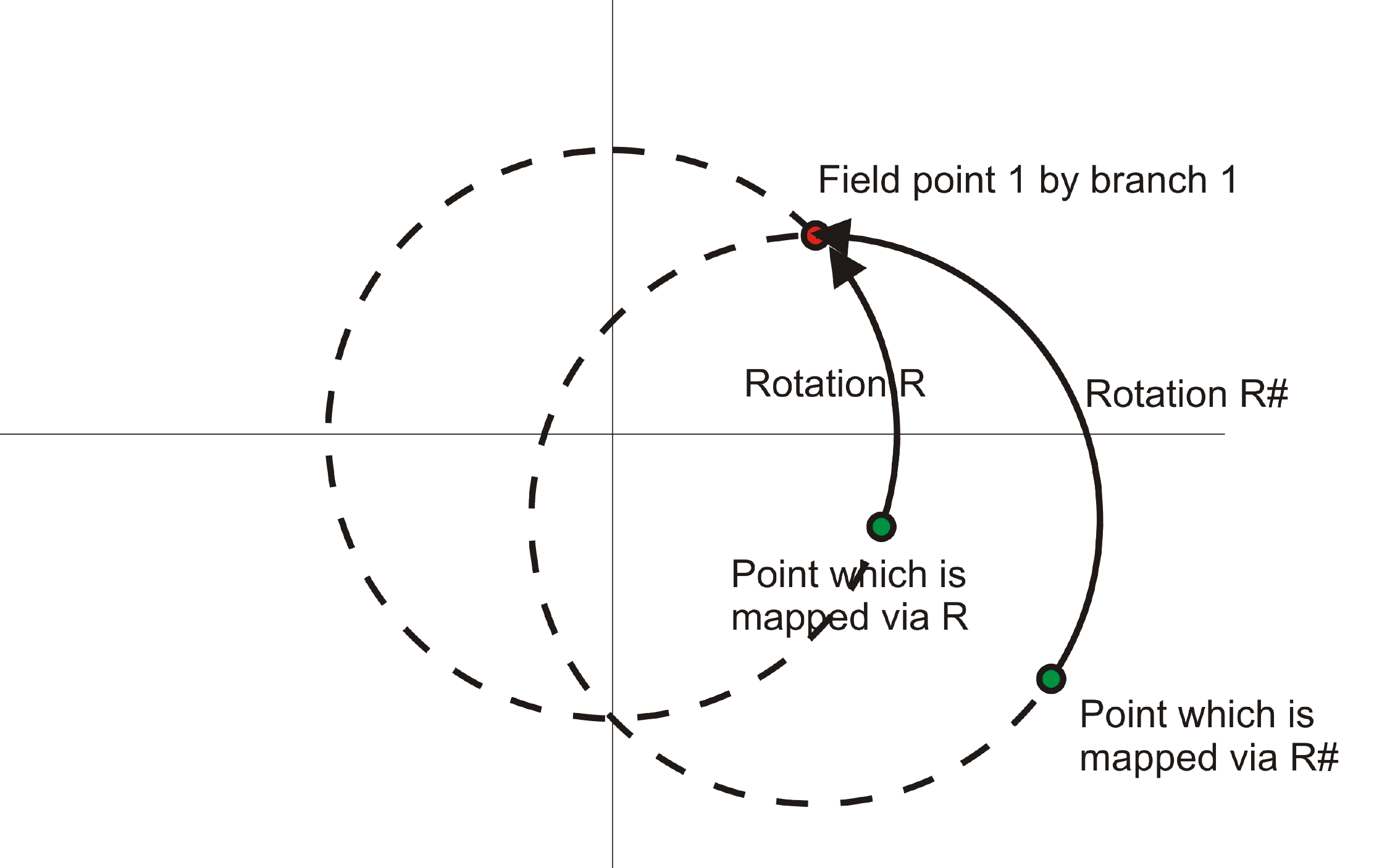}
\caption{\label{x_ray_imag_rot}Image point rotation for a X-ray
  HOLOCAM, design according to Figure \ref{x_ray_gracing_holocam}.}
\end{center}
\end{figure}

Figure \ref{x_ray_gracing_holocam} has shown an example of a HOLOCAM
for X-rays. The, to our knowledge, new development of 3D interferometers
permits quite different design concepts of HOLOCAM
interferometers. We recall that the appropriate design of U is of
crucial importance for the successful application of the HOLOCAM
method. The new 3D transport of fields in 3D interferometers is such a
design concept (Fig. \ref{x_ray_gracing_holocam}). To our best
knowledge this type of 3D interferometers is a new design which is
particularly useful for HOLOCAM applications.

The main difference between HOLOCAM applications in the visible
spectrum and applications in the X-ray range is the availability of
refractive materials. X-ray mirrors can usually only be used for
gracing incidence. This can be respected for a HOLOCAM device so that
the mirrors form some kind of gracing incidence waveguide. Two pairs
of mirror denoted in Figure \ref{x_ray_gracing_holocam} as (1a,2a) and
(1b,2b) form such a guiding structure. Since the guide is bent in 3D
space the transported field is rotated. The two pairs have opposite
bending properties. As a consequence, the field in branch 1 is mutually
rotated with respect to branch 2. Such a rotation has a fixed point
which is not moved by the mapping operation. The two branches in
Figure \ref{x_ray_gracing_holocam} have equal length. U is therefore a
point mapping (The field on one point is approximately mapped on
another point, up to interpolation).

Figure \ref{x_ray_imag_rot} shows the motion of some arbitrary point
of the field in the detector plane. The point moves on a circle around
the fixed point ('center of rotation'). Consequently, only
correlations between field points on such concentric circles are
formed in IF (defined by. Eq. \ref{eq:if_1}). Thus, the solution of
Equation \ref{eq:fund_eq2} is not unique. A second determining map is
needed which is possible by conducting a second field transportation
(called '\#') in Figure \ref{x_ray_imag_rot}. The second superposition
of the two fields has a different shear between the two fields and a
different rotation angle. Figure \ref{x_ray_imag_rot} shows that the
joint and multiple operation of both maps links all
points, indeed. Hence, the fundamental equation (Eq. \ref{eq:fund_eq2}) now
uniquely defines the field $E1$, provided both operations are
evaluated.

The HOLOCAM detector in Figure \ref{x_ray_holocam} can be
displaced. This enables us to measure the diffracted signal at very
different locations while the illumination of the sample and the
sample remain fixed in position. 

The relationship between $E1$ and $E2$ expressed by $E2 = U(E1)$ is an
essential basis for the HOLOCAM concept. This might include multiple
scattering or different order contributions to $E2$. Even a general
diffractive element can be used for the generation of $E2$. The only
important point is the existence of a known linear
relationship between $E1$ and $E2$. The requirements for the image
guiding elements in the HOLOCAM are lower than for X-ray
intensity based microscopes~\cite{pearson}.

Hence, the HOLOCAM has all the potential to improve resolution and
simplify X-ray imaging. The hitherto used phase retrieval from pure
intensity based data is no longer needed. The image evaluation can be
directly based on well known scattering formulas~\cite{born-wolf1991}
which are independent of the actually probed sample.

\section{Conclusion}
The HOLOCAM readdresses the 'reference free phase measurement'. It
thereby reduces many disturbing factors known in reference beam based
digital holography, such as environmental noise and mechanical
instability. The drawback is a twisted path in the HOLOCAM. These
expenses for the internals of a HOLOCAM setup are expected to be
compensated by large savings in the overall 'interferometric
machinery'. To give an example the classical holographic microscope
used for quantitative phase contrast or holographic tomography uses a
design where the sample is part of the interferometer. As shown in the
introduction (Section \ref{introduction}) this setup is rather unreliable
since it causes losses in the accuracy of the phase measurement. Using
a HOLOCAM this is no longer the case. Thus, the same measurements can
be done with greater precision.

For demonstration purposes the design example
(Fig. \ref{vis_holocam}) of the HOLOCAM has well separated explicit
branches. The large design freedom allows internal reflection devices
or twisted optical paths defined by polarization. This is for the
moment reserved for subsequent development. The advantage of the
HOLOCAM is the design freedom in the physical setup and the simplicity
in the mathematical evaluation. Both should lead to compact device
solutions.

Even X-ray imaging should benefit from these concepts. The HOLOCAM can
be realized by state of the art grazing incidence mirrors. A lens is
needed for the integration in the measurement system. This might be
considered as a drawback compared to lenseless systems. Yet zone
lenses are nowadays available and aberrations and other lens errors
are not critical for the HOLOCAM since the effects can be corrected
numerically. The HOLOCAM has all the potential to improve measurement
quality in X-ray systems. This should be enough justification for the
introduction of an additional lens to the system.


%

\end{document}